\documentclass[12pt]{article}
\usepackage{graphicx,epsfig,epsf,amssymb}
\usepackage{scicite}

\usepackage{times}

\def\gsim{\;\lower4pt\hbox{${\buildrel\displaystyle >\over\sim}$}\;}
\def\lsim{\;\lower4pt\hbox{${\buildrel\displaystyle <\over\sim}$}\;}
\def\grls{\;\lower4pt\hbox{${\buildrel\displaystyle >\over <}$}\;}

\newenvironment{sciabstract}{%
\begin{quote} \bf}
{\end{quote}}

\def \<{\langle}
\def \>{\rangle}

\newcounter{lastnote}

\newcommand{\degree}{^\circ}

\title{The Origin of the WMAP Quadrupole}
\author
{Hao Liu$^{1\ast}$, Shao-Lin Xiong$^{1,4}$ and Ti-Pei Li$^{1,2,3\ast}$\\
\normalsize{$^\ast$To whom correspondence should be addressed. E-mail: liuhao@ihep.ac.cn;
litp@tsinghua.edu.cn}}

\date{}

\begin{document}

\baselineskip24pt

\maketitle

{\footnotesize
\begin{enumerate}
\item{Particle Astrophys. Lab., Inst. of High Energy Phys., Chinese Academy
of Sciences, Beijing, China}
\item{Department of Physics \& Center for Astrophysics, Tsinghua University, Beijing, China}
\item{Department of Engineering Physics \& Center for Astrophysics, Tsinghua University, Beijing, China}
\item{Graduate School of Chinese Academy of Sciences, Beijing, China}
\end{enumerate}
}

\begin{sciabstract}
The cosmic microwave background (CMB) temperature maps from the Wilkinson Microwave Anisotropy Probe (WMAP) are of great importance for cosmology. In previous work we had developed a pipeline for map-making
independently of the WMAP team. The new maps produced from the WMAP raw data by our pipeline are notably different to the official ones, and the power spectrum as well as the best-fit cosmological parameters are significantly different too. What's more, by revealing the inconsistency between the WMAP raw data 
and their official map, we had pointed out that there must exist an unexpected problem in the WMAP team's pipeline. In this work, we find that the trouble comes from the inaccuracy of antenna pointing direction caused by a systematical time drift between the attitude data and the science data in the WMAP raw time-order data (TOD). The CMB quadrupole in the WMAP release can be exactly generated from a differential dipole field which is completely determined by the spacecraft velocity and the antenna directions without using any CMB signal. After correcting the WMAP team's error, the CMB quadrupole component disappears. Therefore, the released WMAP CMB quadrupole is almost completely artificial and the real quadrupole of the CMB anisotropy should be near zero. Our finding is important for understanding the early universe.
\end{sciabstract}

The WMAP mission makes measurements of temperature with two antennas, A and B, and record in time-order the difference between the two antenna temperatures, $T_A-T_B$, which is called the time-order data (TOD)~\cite{ben03}. Recently, we have found a serious inconsistency between the WMAP raw data and their 
official map, which indicates that there must exist an unexpected problem in the WMAP team's pipeline for map-making~\cite{liu10}. The WMAP team uses quaternion to describe the attitude of the spacecraft~\cite{hin03,lim}. The spacecraft attitude and the line-of-sight data of the antennas in spacecraft coordinate are used together to determine the antenna directions. In order to reduce the data size, roughly only one quaternion is recorded for each WMAP science frame, whereas each science frame contains 12-30 observations.
Therefore, the quaternion data must be interpolated to determine the spacecraft attitude for each observation. To do the interpolating, the offset of each observation to the start of the frame must be used. For example, for the Q-band, there are 15 observations in one frame. Relative to the time separation between two successive observations, we use $0.000, 0.066, \cdots, 0.933$ as fractional offsets for observation $1, 2, \cdots, 15$ respectively, but the WMAP team take $0.033, 0.099, \cdots, 0.967$ as the offsets. The difference between the resulting antenna directions from different interpolation offset settings is small, only $\sim7'$, just about a half-pixel in the WMAP resolution (with HEALPix resolution parameter $N_{side}=512$~\cite{gor05}).

The observed CMB signal is contaminated by Doppler effect induced by the joint motion of the solar system and the spacecraft. The aroused differential dipole signal
\begin{equation}
d = \frac{T_0}{c}\, \bf{V} \cdot (\bf{n_{_A}} - \bf{n_{_B}})
\end{equation}
with $T_0=2.725$\,K being the CMB monopole, $c$ the speed of light, $\bf{V}$ the joint velocity, $\bf{n_{_A}}$ and $\bf{n_{_B}}$ the unit direction vectors of the two antennas. Although the direction deviation between the two offsettings is small, it can cause the differential dipole signal calculated from Eq.~(1) to be biased  by 10-20\,$\mu$K, which cannot be ignored in comparing to the very weak CMB signal.

\begin{figure}[t]
%\vspace{-1cm}
\begin{center}
\includegraphics[height=5cm, angle=90]{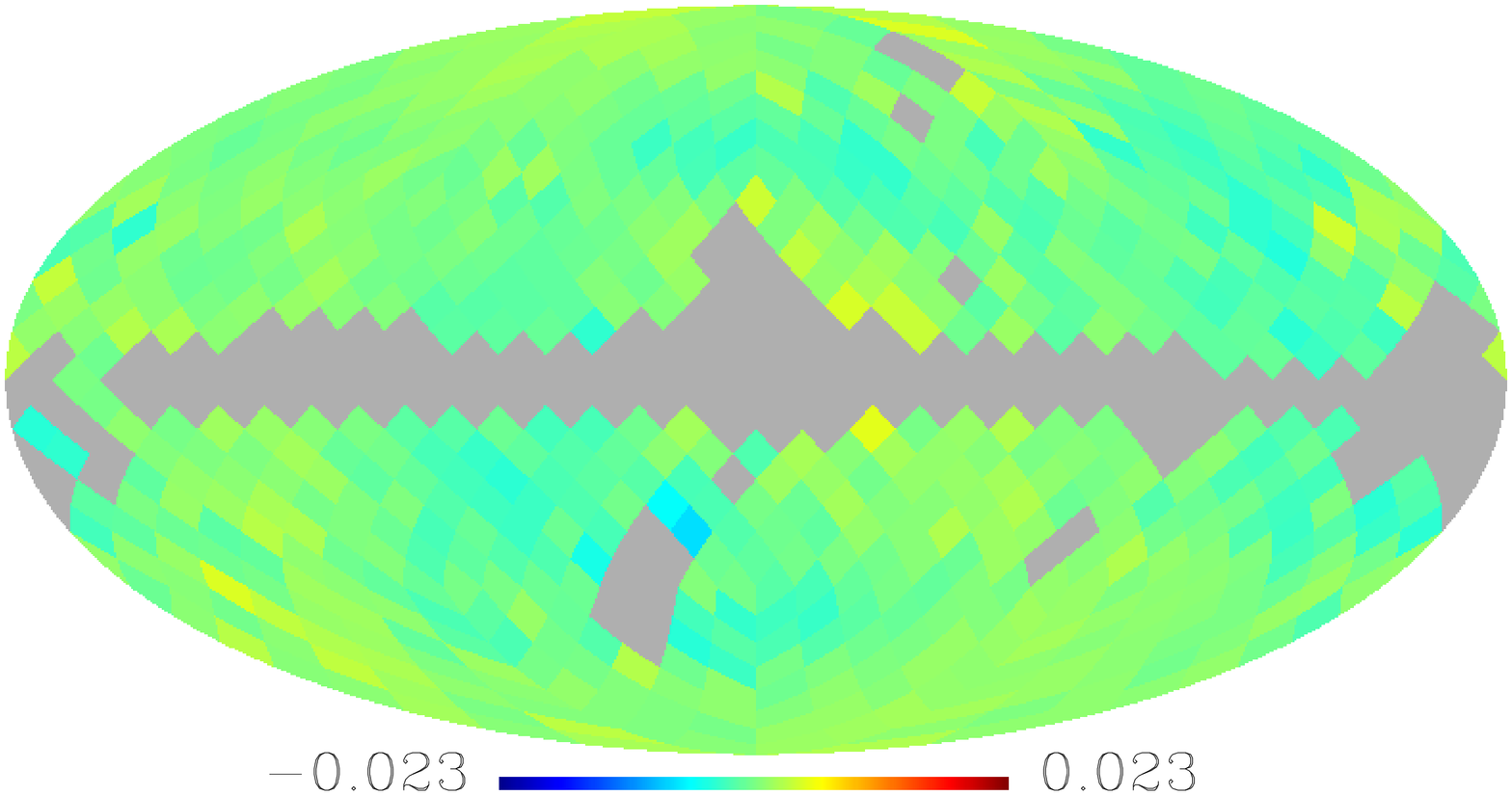}
\includegraphics[height=5cm, angle=90]{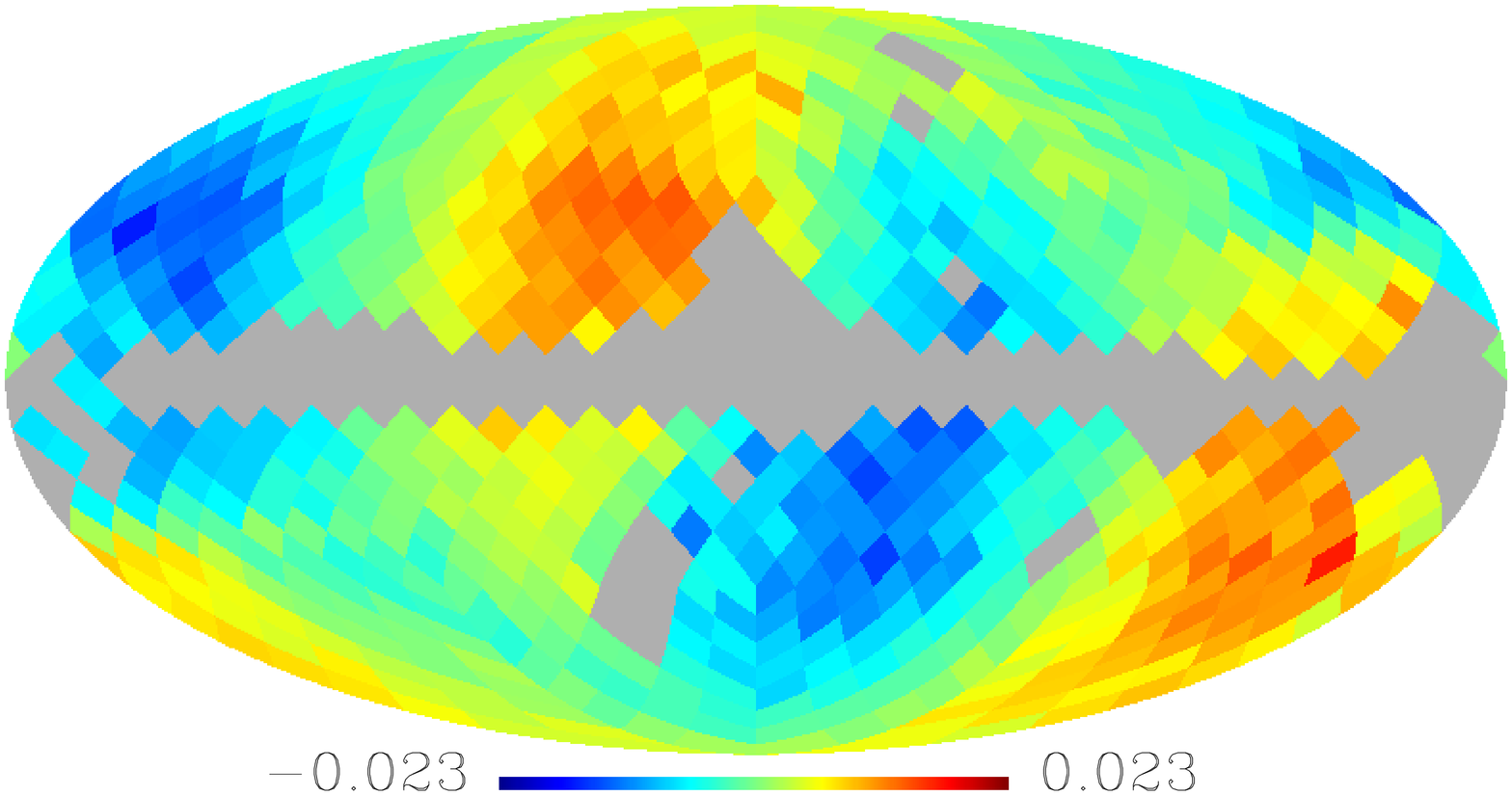}
\end{center}
%\vspace{-7mm}
\caption{ \footnotesize{The difference between the released WMAP3 year-1 map and
our map produced from the same WMAP Q1-band data,
in Galactic coordinates and in units of mK.
{\sl Left panel}: our pipeline with the WMAP quaternion interpolating offset setting.
{\sl Right panel}: our pipeline with our interpolating offset setting.
}}\label{fig:diff_we_wmap}
\end{figure}

Using our pipeline~\cite{lsoft}, we produce CMB maps from the WMAP3 year-1 TOD of Q1-band in our offset setting mode and the WMAP mode, respectively. Figure~\ref{fig:diff_we_wmap} shows the differences between
the WMAP release and our results. The left panel of Fig.~\ref{fig:diff_we_wmap} shows that the result from our softwares using the WMAP offset setting is very close to the WMAP release, indicating that our pipeline is perfectly consistent to the WMAP pipeline only except for the quaternion interpolation offset setting, and the right panel indicates there exists a serious effect of the quaternion interpolation offset setting on the resulting map.

\begin{figure}[t]
%\vspace{-1cm}
\begin{center}
\includegraphics[height=5cm, angle=90]{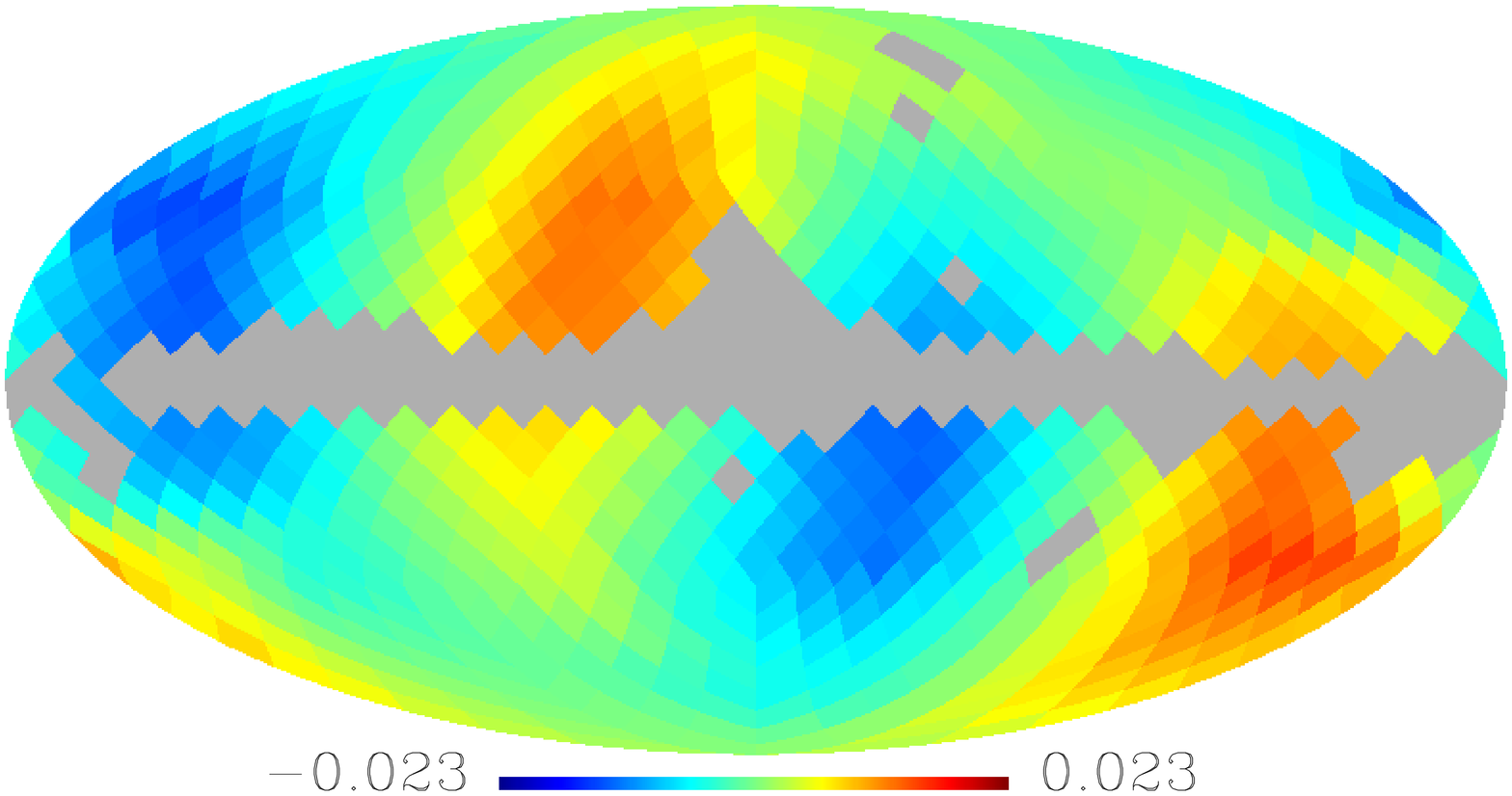}
\includegraphics[height=5cm, angle=90]{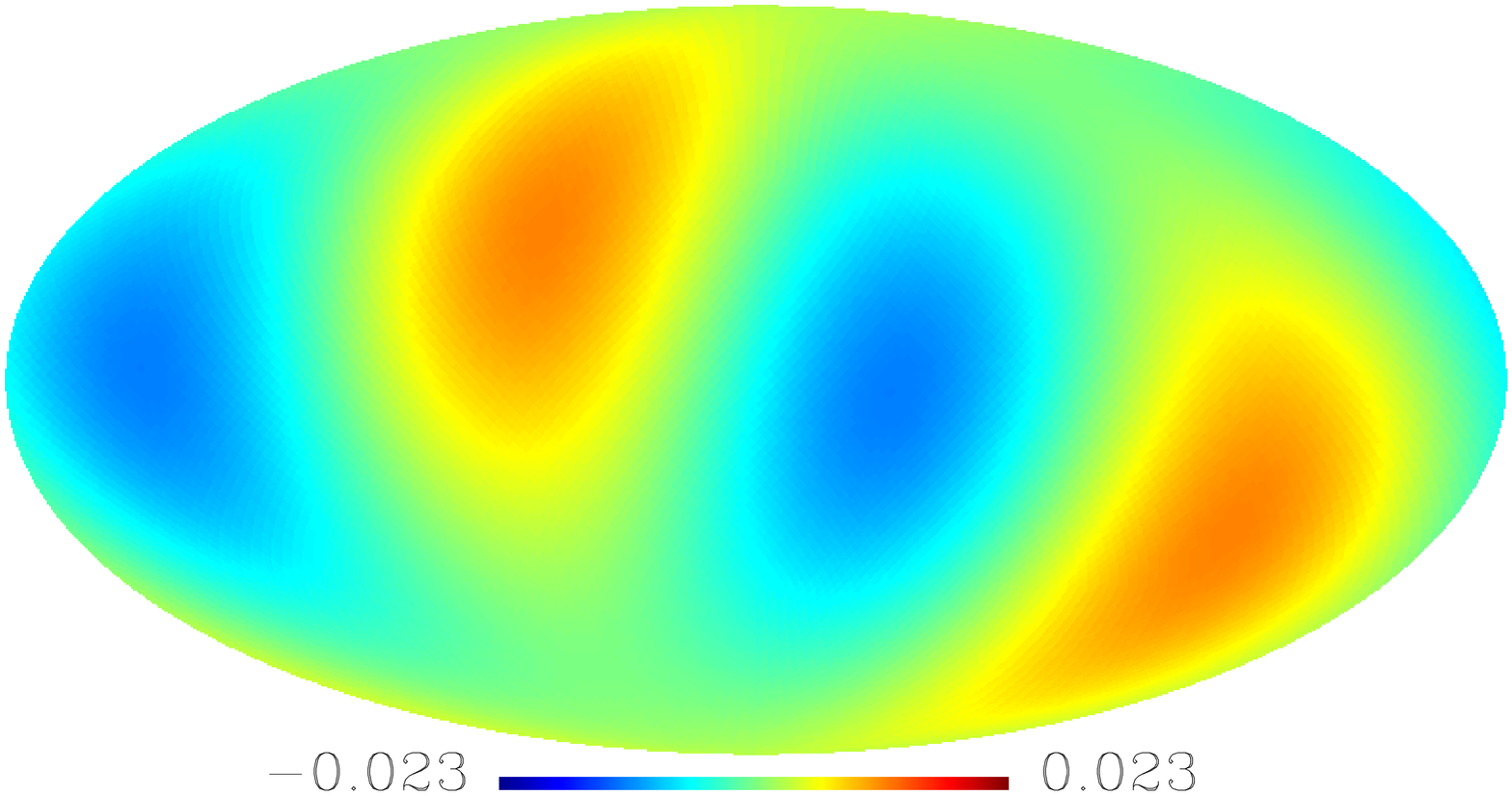}
\end{center}
%\vspace{-7mm}
\caption{ \footnotesize{ {\sl Left panel}:The temperature map produced from the differential dipole field
$d\,'$.  {\sl Right panel}: The released WMAP CMB quadrupole component derived from WMAP5 V and W bands. Both in Galactic coordinates and in units of mK.
}}\label{fig:artificial_quadrupole}
\end{figure}

The differential dipole has to be removed from the raw data to generate usable TOD for map-making, thus dipole deviation will distort the used TOD as well as the resulting temperature map. The deviation of differential dipole
\begin{equation}
d\,'=d_{_{LL}}-d_{_{WMAP}}\,
\end{equation}
where $d_{_{LL}}$ and $d_{_{WMAP}}$ denote the differential dipoles derived from Eq.~(1) with our offset setting mode and the WMAP mode, respectively. We now use the differential dipole field $d\,'$ caused by the
deviation of antenna direction to produce a temperature map. The result is shown in the left panel of Fig.~\ref{fig:artificial_quadrupole}. The right hand-side plot of Fig.~\ref{fig:artificial_quadrupole} is the quadrupole component of the CMB map released by the WMAP team. Comparing the two plots of Fig.~\ref{fig:artificial_quadrupole}, we see that the temperature distortion from the deviating field $d\,'$, which is completely determined by the inaccuracy of antenna direction, is very well consistent with the released
WMAP quadrupole component.

It has to be emphasized that only the spacecraft attitude information is used to compute $d\,'$. Therefore, the left hand-side of Fig.~\ref{fig:artificial_quadrupole} is produced without any CMB information at all. On the contrary, the WMAP quadrupole is released by the WMAP team to describe the characteristic of the CMB anisotropy. That the map recovered from $d\,'$ is identical with the WMAP quadrupole certainly indicates that the released CMB quadrupole is completely artificial, and the real quadrupole signal observed by the WMAP spacecraft should be near zero. After correcting the mistake in the WMAP routine, the false quadrupole component really disappears almost completely, as shown by the identity between the two plots of Fig.~\ref{fig:artificial_quadrupole}.

The WMAP spacecraft is continuously observing; therefore, if the quaternion datum and the corresponding differential datum are recorded simultaneously, the WMAP setting that uses 0.5 as the first offset will be correct. However, we find that for all bands the quaternion datum is recorded $\sim25.6$ ms later than the corresponding differential datum\footnote{To check for the asynchronous, one can use the fv tool to open any fits format WMAP TOD file, and check the first time record in the Meta Data Table (contains the quaternion) and the Science Data Table (contains the differential data). The difference between the two times is about $2.963\times 10^{-7}$ in Julian time format (1.0 for one day), with the quaternion time being later than the differential datum time. Since the time record in the Science Data Table is used for all bands, the time drift is the same for all bands.}. To obtain a correct quaternion for the corresponding differential data, one must extrapolate it to an earlier time and it is just what we do in our pipeline. Therefore, our result should be more reliable than the WMAP release.

\begin{figure}[i]
\begin{center}
\includegraphics[height=0.3\textwidth, angle=90]{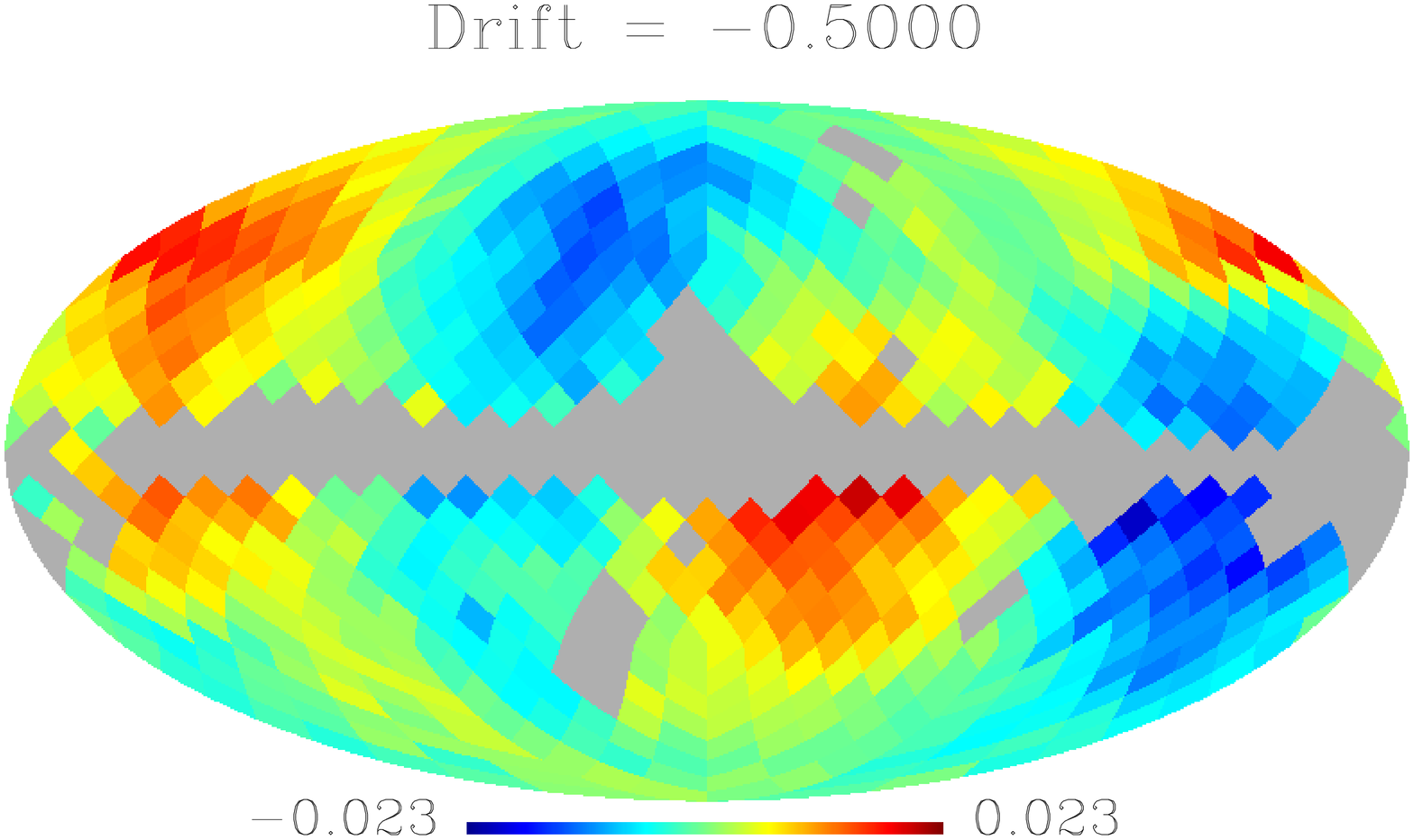}
\includegraphics[height=0.3\textwidth, angle=90]{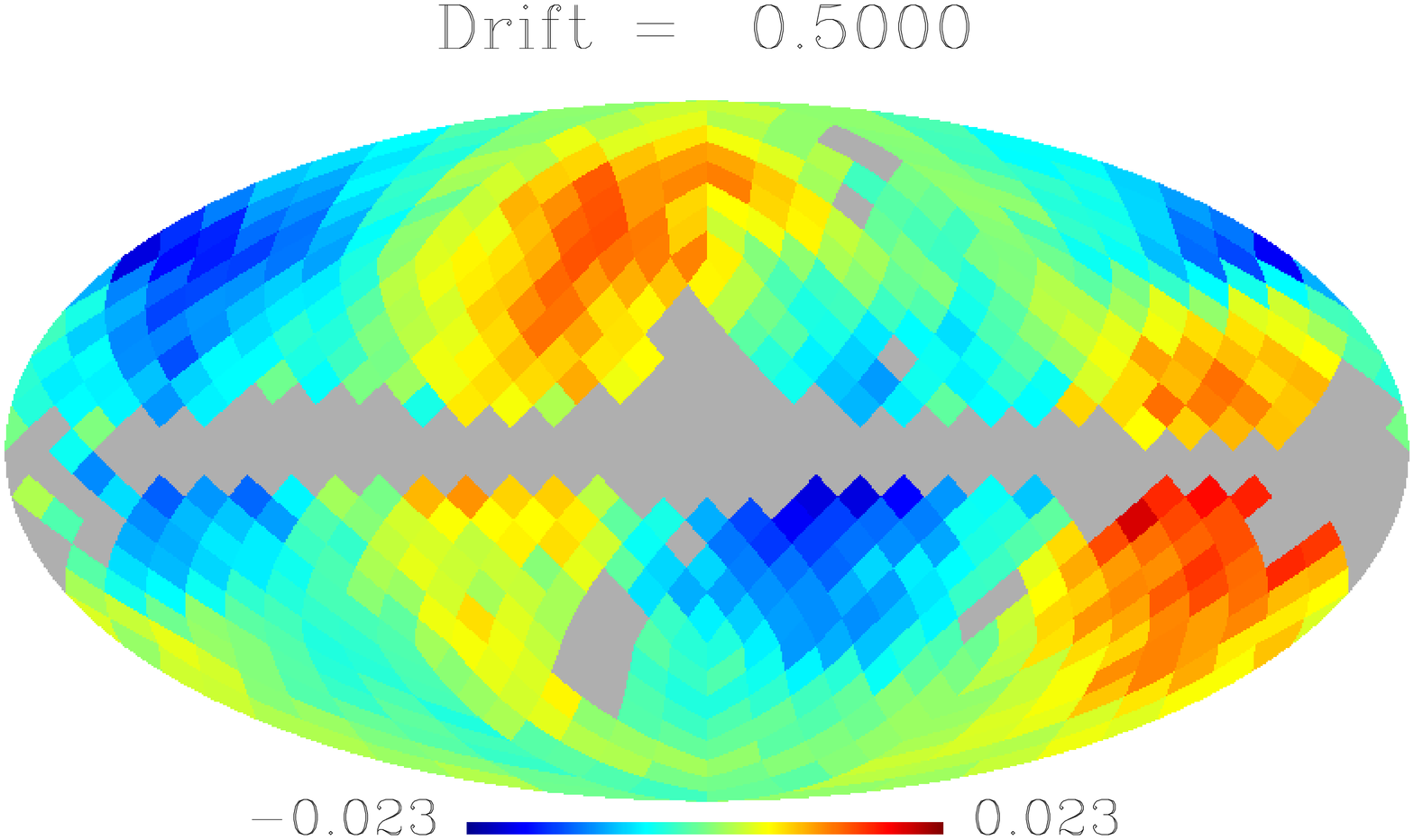}

\includegraphics[height=0.3\textwidth, angle=90]{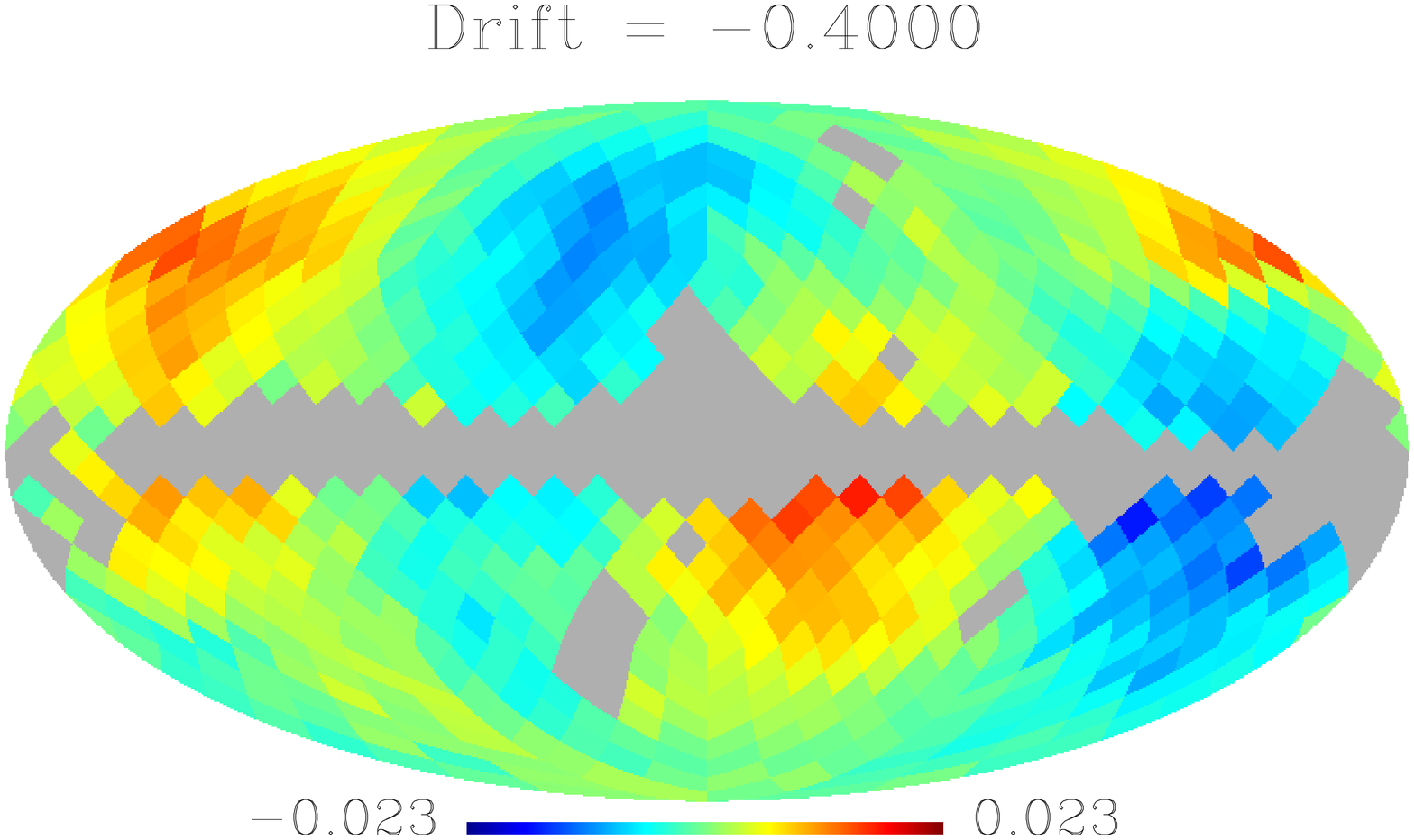}
\includegraphics[height=0.3\textwidth, angle=90]{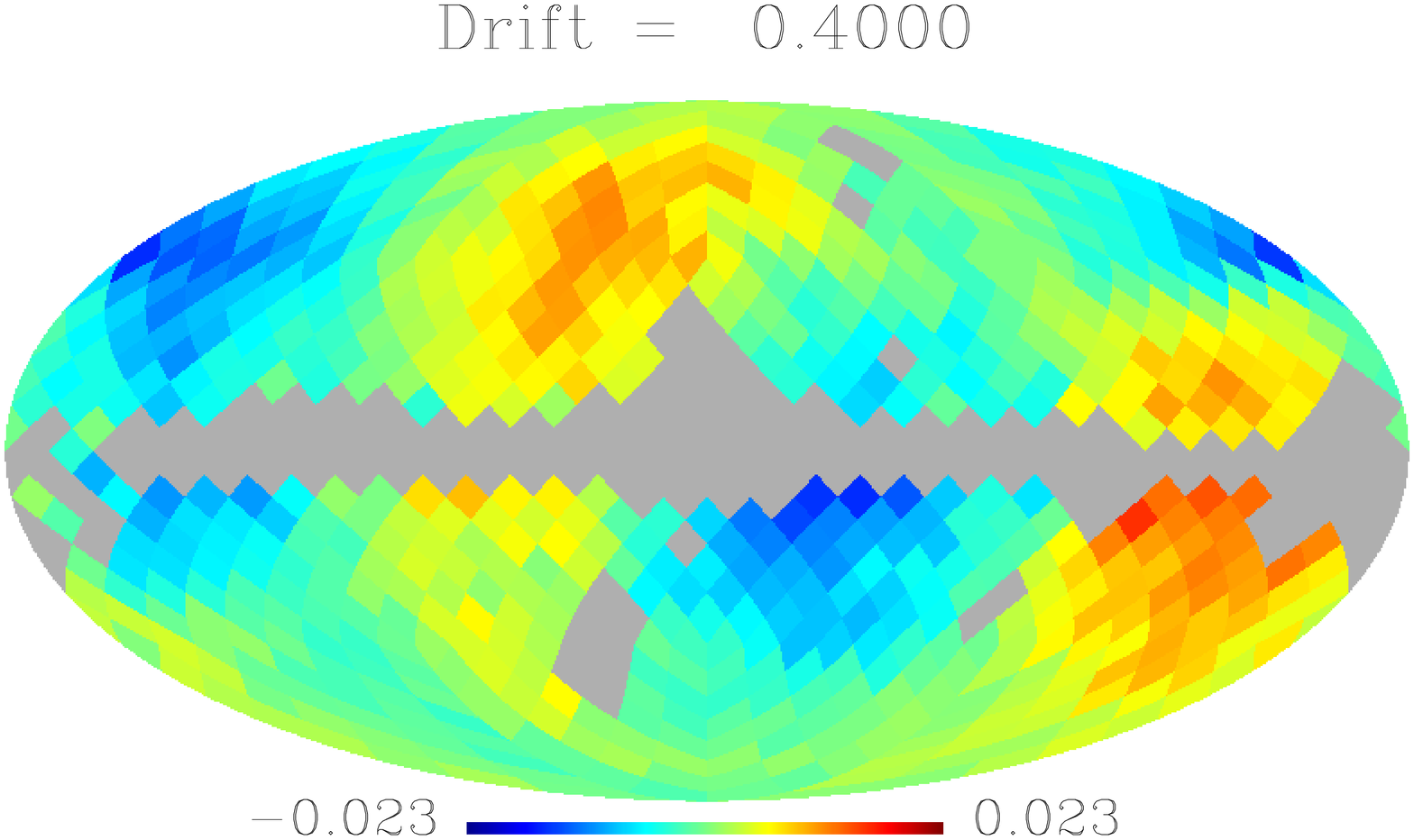}

\includegraphics[height=0.3\textwidth, angle=90]{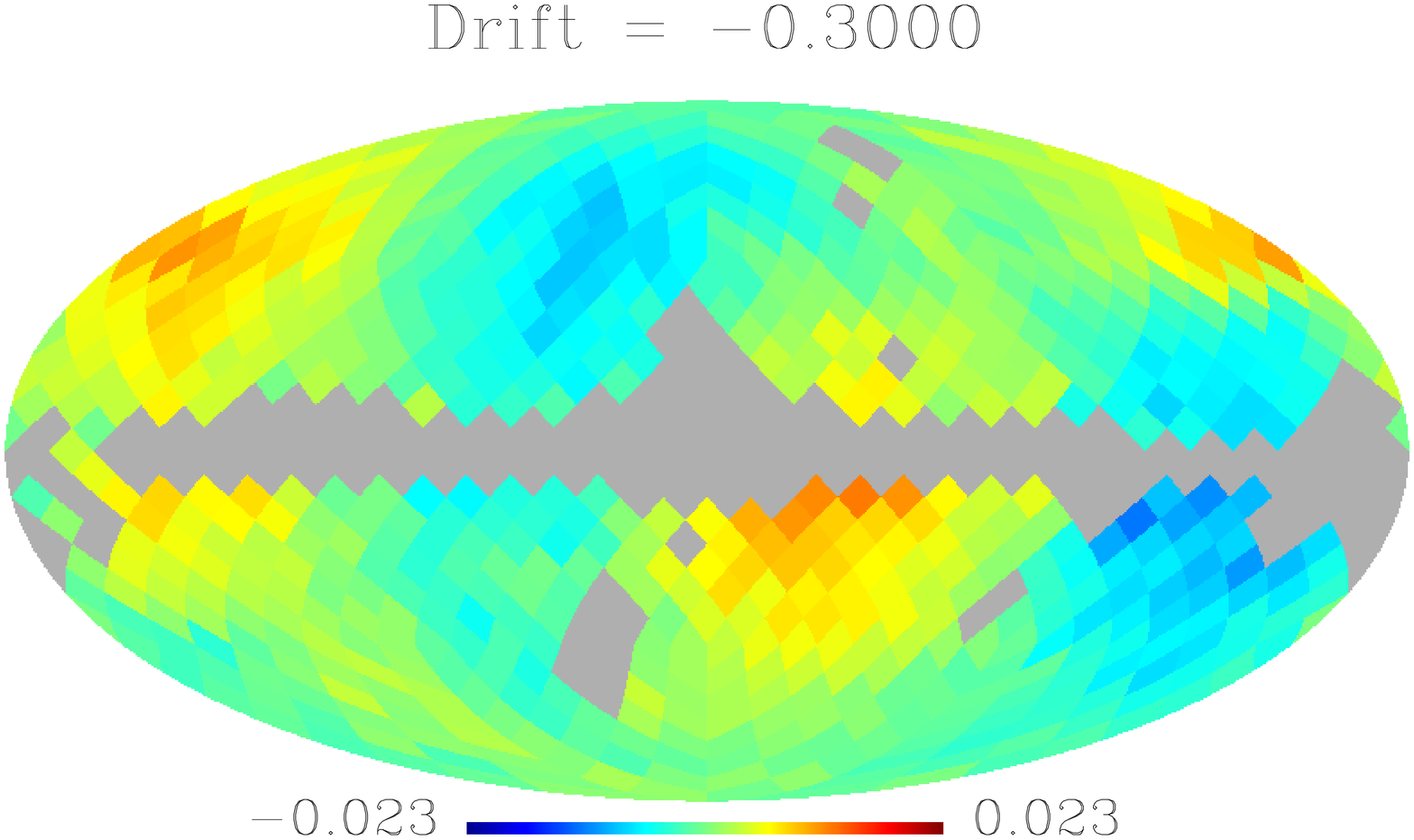}
\includegraphics[height=0.3\textwidth, angle=90]{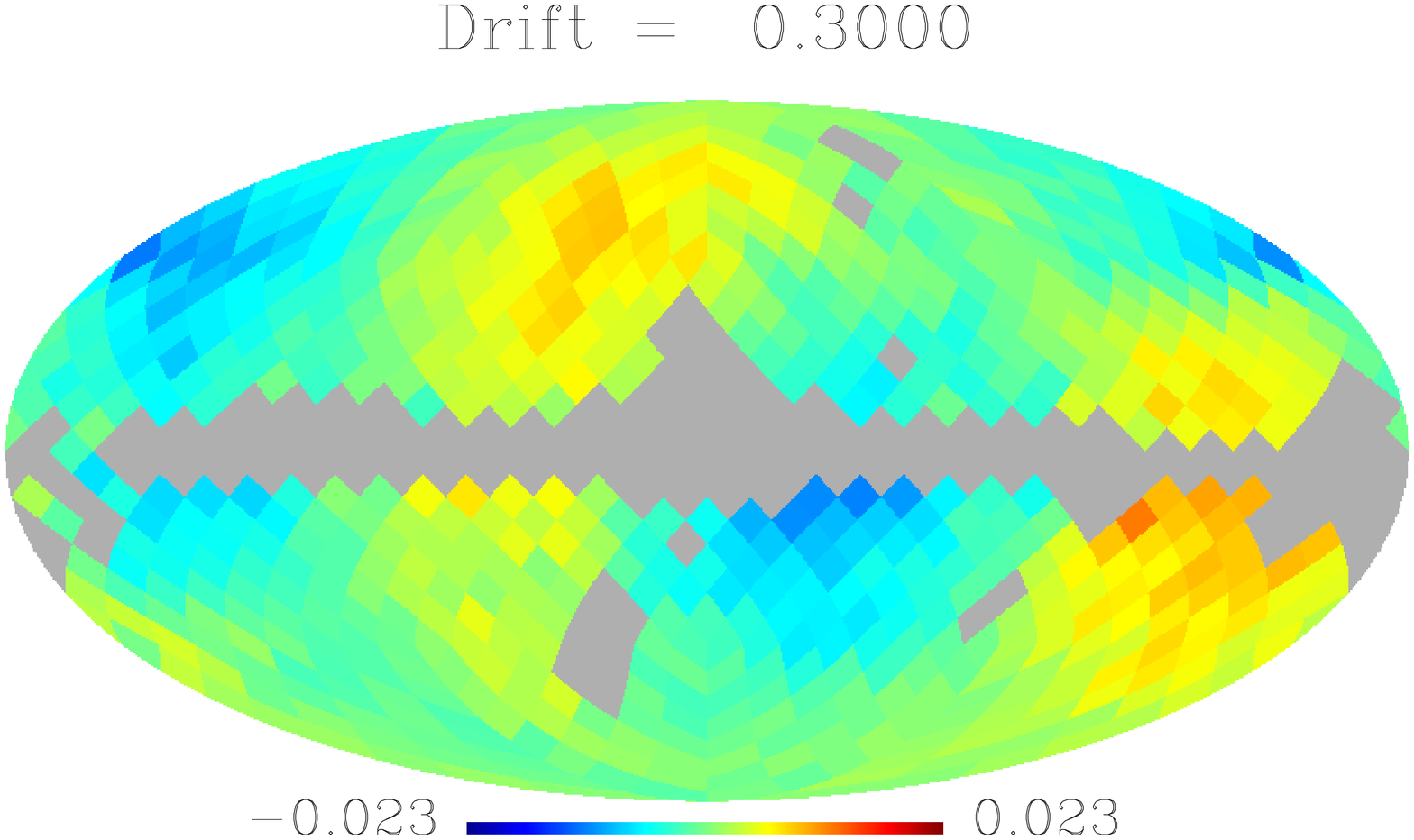}

\includegraphics[height=0.3\textwidth, angle=90]{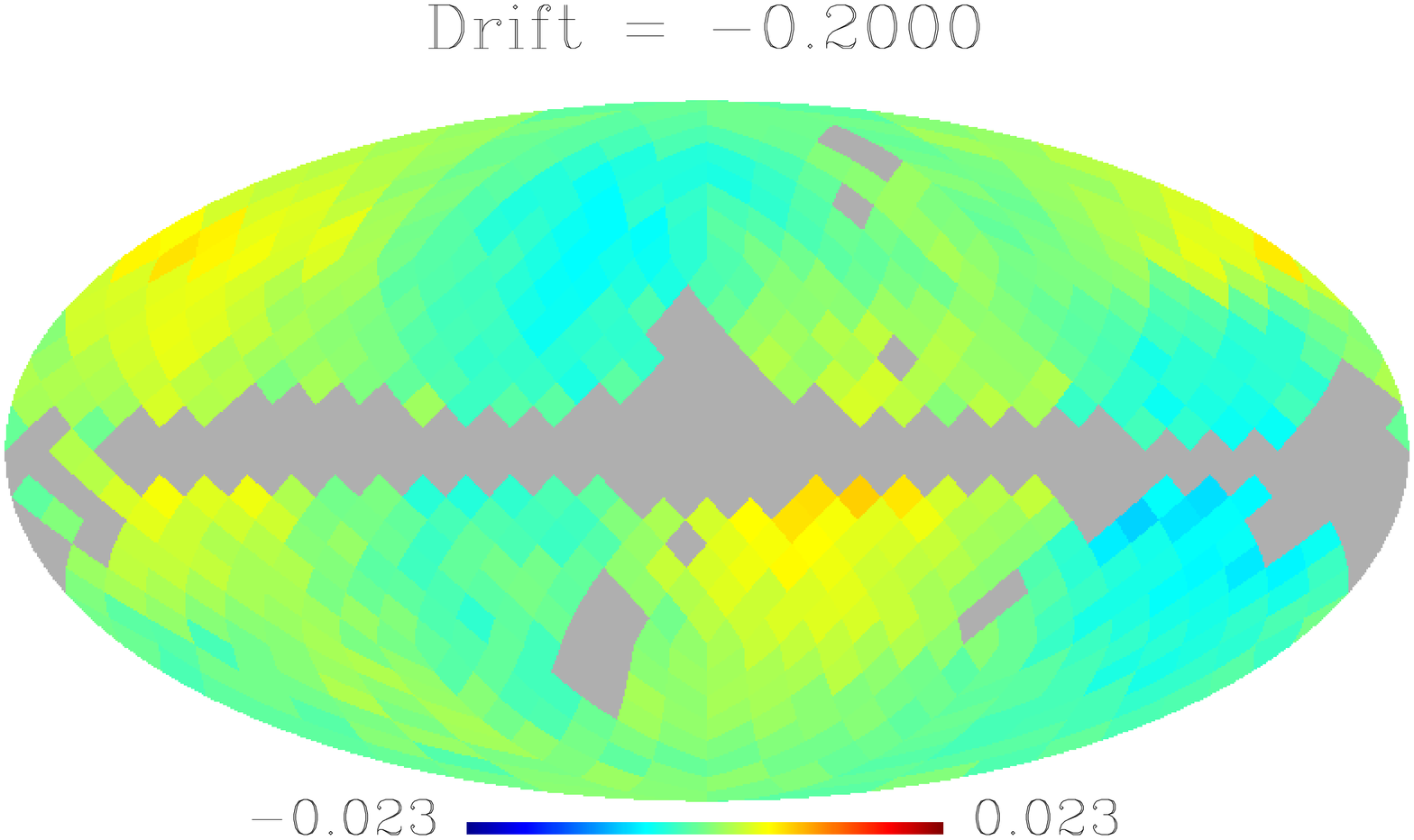}
\includegraphics[height=0.3\textwidth, angle=90]{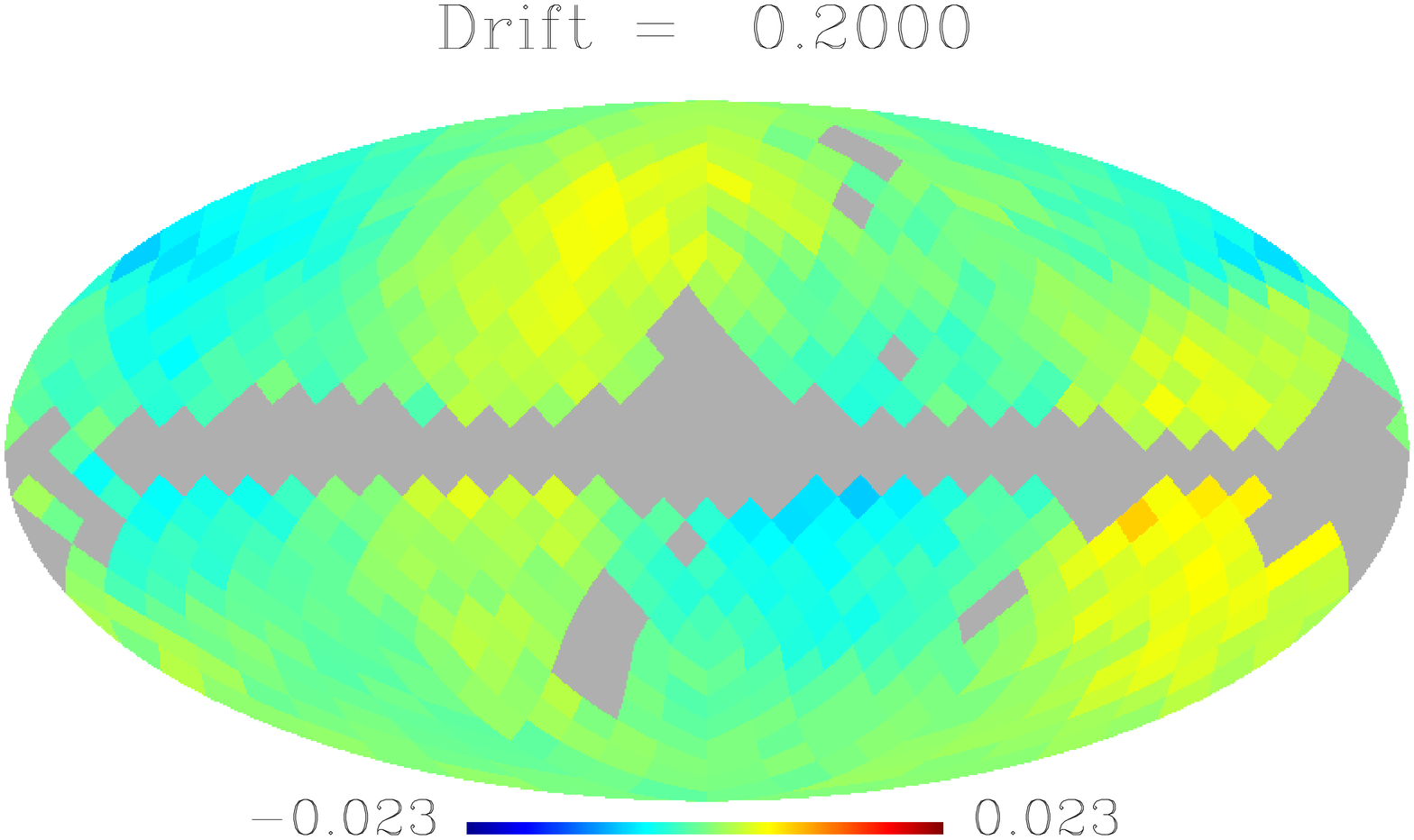}

\includegraphics[height=0.3\textwidth, angle=90]{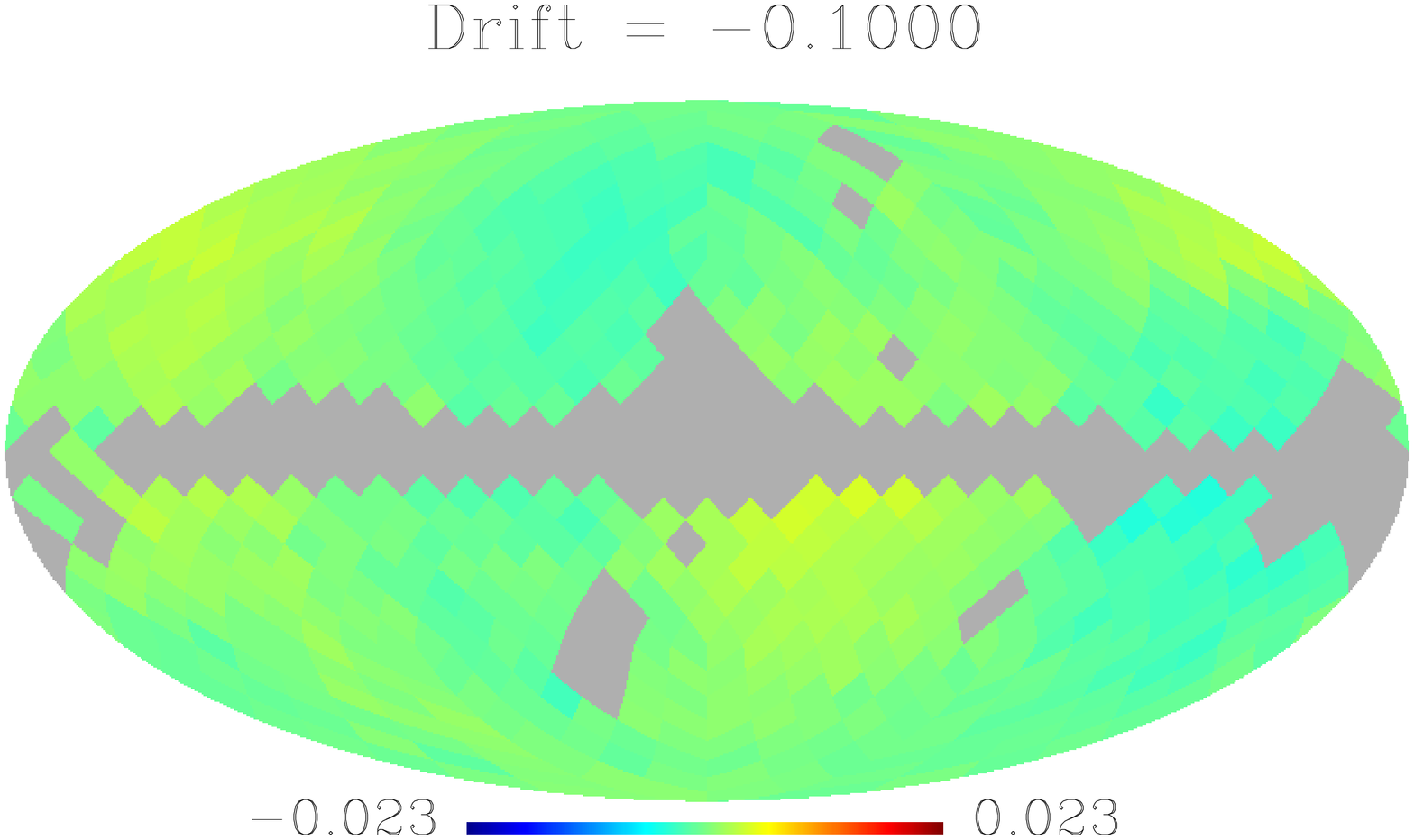}
\includegraphics[height=0.3\textwidth, angle=90]{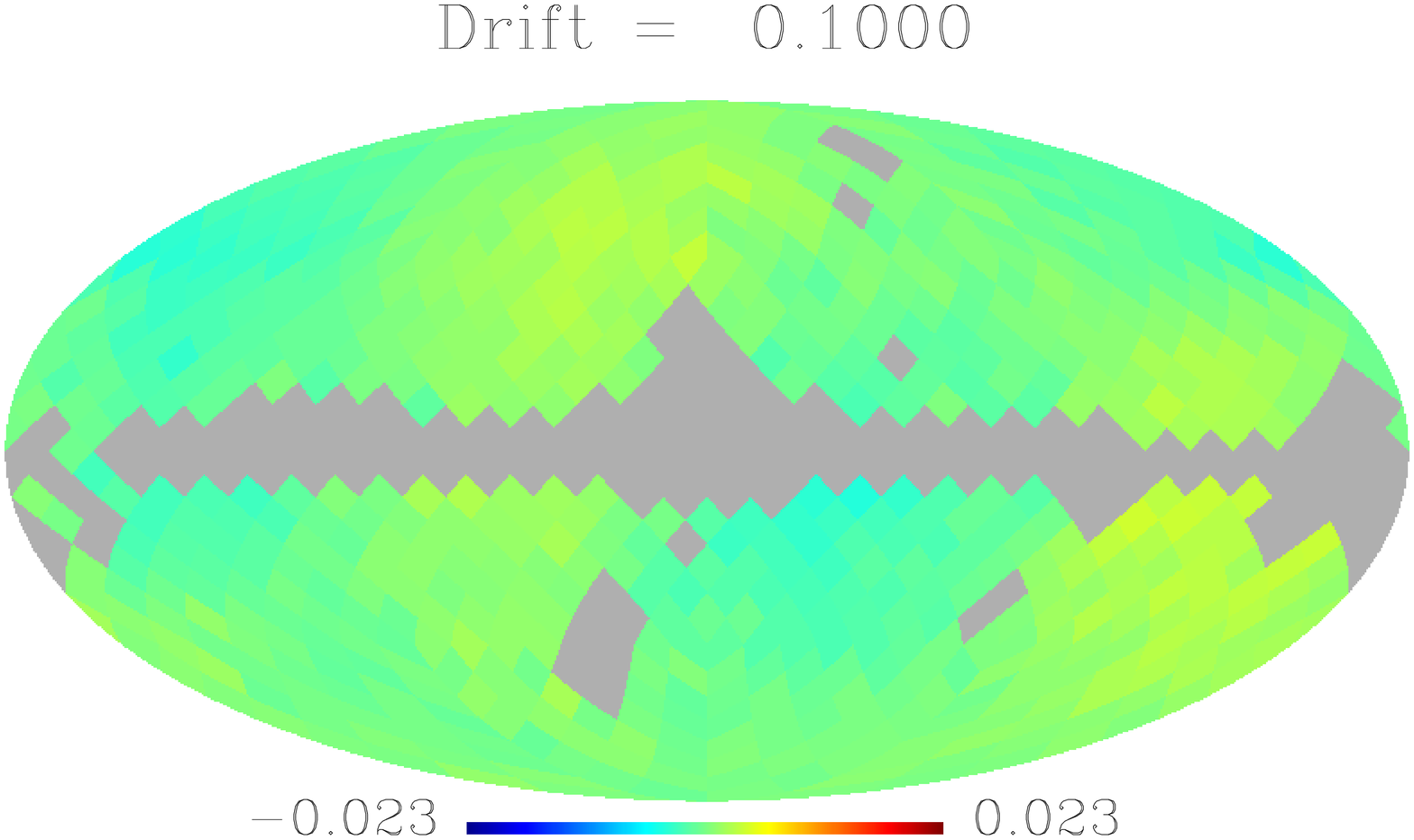}
\end{center}
\caption{ \footnotesize{ The residual maps for different offsetting with downgraded resolution
of $N_{side}=8$, in Galactic coordinates and in units of mK. From top to bottom the absolute first-offsets: 0.1, 0.2, 0.3, 0.4, 0.5. {\sl Left panel}: Negative first offsets. {\sl Right panel}: Positive first offsets.
}}\label{fig:drifts}
\end{figure}

It's now clear that the inconsistency we found~\cite{liu10} between the WMAP TOD and released maps originates from the inconsistency between settings in their data base and pipeline. It has to be noted that this inconsistency is not necessary to produce a systematic deviation of the position of a bright radio source in a map recovered from the TOD, because for a single observation the direction deviation caused by the time drift is dependent with the scan path in the science frame. Since the systematical time drift between the attitude data and the science data in the WMAP TOD can affect the entire calibration and map-making, there might be more unexpected consequences. At the present time, other unexpected problems existed still in the WMAP system cannot be completely excluded too. Therefore, the WMAP TOD and pipeline have to be thoroughly checked and corrected together by the WMAP team, and a new consistent ones should be then published for cosmology study. At the meantime, to find a proper offset we produce maps from the WMAP3 Q1-band year-1 TOD for a range of offsets, and calculate the rms amplitude for each offset. As we stated earlier, the direction deviation caused by a wrong offset should in general create additional noise in the resulting map and increase the RMS. The offsetting with minimum RMS should be taken as a proper one, unless the contamination happens to anti-correlated by chance with the cosmological signal. Fig.~\ref{fig:drifts} shows the maps produced with 11 different first offsets: $0.0, \pm0.1, \pm0.2, \pm0.3, \pm0.4, \pm0.5$. The maps in Fig.~\ref{fig:drifts} are after subtracting the zero first-offset map and down-graded to $N_{side}=8$, their temperature RMS are given in Table~1. From Table~1 one can see that our offsetting with the zero first-offset is really suitable to be used to compensate the effect of the time drift between the quaternion and the differential data. 

We can see clearly from Fig.~\ref{fig:drifts} that the WMAP-like quadrupole components are very well proportional to the first offset, showing a typical picture of an artificial origin and hardly to be contributed by the true CMB. It is important to notice from Fig.~\ref{fig:drifts} and Table~1 that, the quadrupole components in the positive and negative offset maps are almost exactly symmetric, it further ensures that the true CMB quadrupole power should be exactly very close to zero. Although the absence of CMB quadrupole has been explained by cosmic variance, it is no longer convincing to appeal to cosmic variance as reason for the complete absence of a CMB quadrupole: It is actually impossible for a CMB power spectrum to be negative; therefore, only when the CMB power is exactly counteracted by cosmic variance can we see a zero quadrupole in our inflation picture. 

\begin{table}{Table 1: Temperature RMS of the maps in Fig.~\ref{fig:drifts}}\\[1ex]
\begin{tabular}{cccc}  \hline
First-offset &\vline& \multicolumn{2}{c}{RMS ($\mu$K)}\\
magnitude &\vline& Negative offset &Positive offset \\
\hline
0.1& \vline& 1.49 &1.49   \\
0.2 & \vline& 2.92 &2.93   \\
0.3& \vline& 4.38 &4.35   \\
0.4& \vline& 5.80 &5.79    \\
0.5& \vline& 7.25 &7.23    \\
\hline
\end{tabular}
\end{table}

It is a surprise that small errors of antenna pointing directions (only about a half-pixel in the WMAP resolution) can lead to serious effects on the official WMAP sky maps. The reason is that the directions are not only used to construct the sky map, but also used to calculate the dipoles for each observation, which are generated by the relative motion of the spacecraft to the CMB and have to be removed from the raw data before map-making because their intensities roughly 10 to 20 times greater than those of the CMB anisotropies.The error of antenna direction will produce an error in predicted dipole intensity which can no longer be ignored. As a common problem the accuracy of direction measurement should be studied more careful in CMB experiments. 

A great achievement of the WMAP mission is to validate the concordance cosmology model $\Lambda$CDM: a flat-$\Lambda$ dominated universe with initial nearly scale invariant adiabatic Gaussian fluctuations. After correcting the inconsistency between the WMAP raw data and recovered map, we find~\cite{liu09_1} that the acoustic peaks of the CMB angular power spectrum are better consistent with the other CMB experiment BOOMERANGE, and the best fit cosmological parameters match with improved accuracy those of other two independent experiments: the distance measurements from the Type Ia supernovae and the baryon acoustic oscillations in the distribution of galaxies. Therefore, finding and correcting the error in the WMAP routine is not a catastrophe for WMAP, just on the contrary, it makes the WMAP results to validate the concordance cosmology with improved accuracy.

The near-vanishing of the CMB quadrupole apparently seems to be in contradiction to the expectation of scale invariance in the primordial fluctuations. However, the absence of large-angular power in CMB temperatures is not necessarily to be opposed to the concordance model. From the inflation model, the CMB anisotropy at the largest angular scales is of direct probe of the primordial density perturbations during the inflation. The prediction of a flat power spectrum of the CMB at large-angular scales from the standard model is based on two presuppositions: 1) The primordial perturbation spectrum is of Harrison-Zeldovich type; 2) For all angular scales we measure the amplitudes of the fluctuations on them today, the times when they first left the horizon during the inflationary era were all after the primordial perturbations production. The CMB quadrupole amplitude is a measurement of the magnitude of CMB fluctuation
at the angular scale of $90\degree$. If the time when the angular scale of $90\degree$ left the horizon was as early as before the reheating period while the primordial density perturbations had not occurred,
we could detect no CMB quadrupole. Thus the absence of CMB quadrupole may indicate that we have seen
a rigid universe in a supercooled state at the very early period which we can explore through low-$l$ (large scale) cosmology, such a version should excite every physicists.

The authors greatly thank Prof.\,Peter Freeman for providing his work~\cite{Freeman06} and helping us to cross-check our map-making pipeline, Charling Tao and Boud Roukema for helpful discussions and suggestions. This work is supported by the National Basic Research Program of China (Grant No. 2009CB-824800), the CAS project KJCX2-YW-T03 and China Postdoctoral Science Foundation funded project H91I21734A. The data analysis made use of the WMAP data archive and the HEALPix software packages.

\bibliography{scibib}
\bibliographystyle{Science}

%\begin{scilastnote} \item \end{scilastnote}

\clearpage

\end{document}